\documentstyle[twocolumn,aps,epsf]{revtex}
\begin{document}
\draft
\twocolumn[\hsize\textwidth\columnwidth\hsize\csname
@twocolumnfalse\endcsname

\title{Orbital Fluctuation-Induced Triplet Superconductivity :\\
Mechanism of Superconductivity in ${\rm Sr}_{2}{\rm RuO}_{4}$
}

\author{Tetsuya Takimoto}
\address{Electrotechnical Laboratory, Tsukuba, Ibaraki 305}

\date{Received 22 June 2000}
\maketitle
\begin{abstract}
The mechanism of superconductivity in ${\rm Sr}_{2}{\rm RuO}_{4}$ 
is studied using a degenerate Hubbard model within the weak coupling 
theory. When the system approaches the orbital instability 
which is realized due to increasing the on-site Coulomb interaction 
between the electrons in the different orbitals, 
it is shown that the triplet superconductivity appears. 
This superconducting mechanism is only available 
in orbitally degenerate systems with multiple Fermi surfaces.
\end{abstract}

\vskip2pc]
\narrowtext


Since the discovery of the superconductivity in 
${\rm Sr}_{2}{\rm RuO}_{4}$ below $T_{\rm c}$=1.5K 
experimental and theoretical 
investigations on this compound 
have been carried out intensively. \cite{rf:1}
The main features of the results of investigations may be summarized 
as follows; (1) this compound 
has the same crystal structure as the high $T_{\rm c}$ cuprate 
${\rm La}_{2-x}{\rm Sr}_{x}{\rm CuO}_{4}$, 
(2) the conduction electrons exhibit the 2D Fermi liquid property \cite{rf:2}, 
(3) there are three cylindrical $t_{2g}$ Fermi surfaces \cite{rf:3,rf:4}, 
(4) $T_{\rm c}$ is strongly suppressed to less than 1.5K by the non-magnetic 
impurities \cite{rf:5,rf:18}, 
(5) no Hebel-Slichter peak was seen in $1/T_{1}$ \cite{rf:6}. 
(6) The NMR Knight shift \cite{rf:7} and $\mu$-SR \cite{rf:8} experiments 
indicate that 
the spin triplet state breaking the time reversal symmetry is realized. 

In an early stage of investigations, 
it was considered that the symmetry of the superconducting 
order parameter in ${\rm Sr}_{2}{\rm RuO}_{4}$ is of the $p$-wave similar to 
that in $^{3}$He and the pairing is caused by ferromagnetic spin fluctuation 
enhanced by the Hund's coupling \cite{rf:9}. 
From these facts, it was considered that the most 
consistent superconducting order parameter is represented by 
${\bf d}({\bf k})={\hat{z}}(k_{x}+{\rm i}k_{y})$. Recently, some 
experimental results suggest the existence of line nodes in the 
superconducting gap. \cite{rf:10} This result is incompatible with the above 
assumed form of 
${\bf d}({\bf k})$. Also, the inelastic neutron scattering data does 
exhibit the incommensurate 
magnetic response \cite{rf:11} consistent with the nesting vector 
predicted from the band-structure calculation \cite{rf:12}, 
but no discernible response around ${\bf q}={\bf 0}$. 
In view of these experimental results throwing some doubts about 
the $p$-wave scinario, it is desirable to develop a more detailed theory 
from a microscopic point of view. 

In this letter, we investigate the mechanism of the 
superconductivity in ${\rm Sr}_{2}{\rm RuO}_{4}$ based on a 
degenerate Hubbard model having four on-site interaction parameters; 
the Coulomb integral between the same orbital electrons $U$, 
the direct Coulomb integral between the different orbital electrons $U'$, 
the Hund's coupling constant $J$ and the pair-hopping constant $J'$. 

The model Hamiltonian mentioned above is described 
as follows;
\begin{eqnarray}
  H&=&H_{\rm 0}+H_{\rm I} ,\nonumber\\
  H_{\rm 0}&=&\sum_{{\bf k},\sigma}\sum_{\alpha}\epsilon_{{\bf k}\alpha}
  a_{{\bf k}\alpha\sigma}^{\dagger}a_{{\bf k}\alpha\sigma} ,
  \nonumber\\
  H_{\rm I}&=&\sum_{i}[U\sum_{\alpha}
  n_{i\alpha\uparrow}n_{i\alpha\downarrow}
  +J\sum_{\alpha \neq \alpha'}
  {\bf S}_{i\alpha}\cdot{\bf S}_{i\alpha'}
  \nonumber\\
  &&+U'\sum_{\alpha > \alpha'}n_{i\alpha}n_{i\alpha'}
  +J'\sum_{\alpha \neq \alpha'}
  a_{i\alpha\uparrow}^{\dagger}a_{i\alpha'\uparrow}
  a_{i\alpha\downarrow}^{\dagger}a_{i\alpha'\downarrow}]
  \eqnum{1}\label{modelHam}
\end{eqnarray}
where $a_{{\bf k}\alpha\sigma}$ is 
the Fourier transform of the annihilation operator for the 
$d_{\alpha}$-orbital electrons ($\alpha$=$xy$, $yz$, $zx$). 
$\epsilon_{{\bf k}\alpha}$ describes the energy dispersions of 
the tight binding bands measured from the Fermi level as follows; 
$\epsilon_{{\bf k}\alpha}=-\epsilon_{0}-2t_{x}\cos{k_{x}}-2t_{y}\cos{k_{y}}
+4t'\cos{k_{x}}\cos{k_{y}}$. In what follows we choose the same values 
for the parameter sets 
($\epsilon_{0}$, $t_{x}$, $t_{y}$, $t'$) as in Ref.[13], 
(0.50, 0.44, 0.44, -0.14), (0.24, 0.045, 0.31, 0.01), 
(0.24, 0.31, 0.045, 0.01) eV for $d_{xy}$, $d_{yz}$, $d_{zx}$-orbital, 
respectively. 
These bands are occupied by four electrons per site, corresponding to 
a tetra-valent Ru atom. 
We show the results of calculation for the Fermi surfaces and 
the ${\bf q}$-dependence of the static 
susceptibilities $\chi^{(0)}_{\alpha\alpha}({\bf q})$ in Fig. 1(a) and 1(b), 
respectively. 
Each peak of $\chi^{(0)}_{\alpha\alpha}({\bf q})$ corresponds to the 
nesting vector depicted in Fig. 1(a). 

Following the standard prescription \cite{rf:15,rf:16}, 
we can derive the following effective 
interaction Hamiltonian which is rotationally invariant in the 
spin space;
\begin{eqnarray}
  &&H_{\rm I}^{\rm eff}\nonumber\\
  &=&
  \frac{1}{4}
  \sum_{\alpha,\beta}\sum_{\mu\nu\zeta\eta}\sum_{{\bf k},{\bf k'}}
  V_{\alpha\beta}^{\mu\nu\zeta\eta}({\bf k}-{\bf k'})
  a_{{\bf k}\alpha\mu}^{\dagger}a_{-{\bf k}\beta\nu}^{\dagger}
  a_{-{\bf k'}\beta\zeta}a_{{\bf k'}\alpha\eta}\nonumber\\
  &+&
  \frac{1}{4}
  \sum_{\alpha\neq\beta}\sum_{\mu\nu\zeta\eta}\sum_{{\bf k},{\bf k'}}
  \bar{V}_{\alpha\beta}^{\mu\nu\zeta\eta}({\bf k}-{\bf k'})
  a_{{\bf k}\alpha\mu}^{\dagger}a_{-{\bf k}\alpha\nu}^{\dagger}
  a_{-{\bf k'}\beta\zeta}a_{{\bf k'}\beta\eta},\nonumber\\
  &&V_{\alpha\beta}^{\mu\nu\zeta\eta}({\bf q})
  =S_{\alpha\beta}({\bf q})
  {\bf \sigma}_{\mu\eta}\cdot{\bf \sigma}_{\nu\zeta}
  -C_{\alpha\beta}({\bf q})
  \delta_{\mu\eta}\delta_{\nu\zeta},\nonumber\\
  &&\bar{V}_{\alpha\beta}^{\mu\nu\zeta\eta}({\bf q})
  =\bar{S}_{\alpha\beta}({\bf q})
  {\bf \sigma}_{\mu\eta}\cdot{\bf \sigma}_{\nu\zeta}
  -\bar{C}_{\alpha\beta}({\bf q})
  \delta_{\mu\eta}\delta_{\nu\zeta}\eqnum{2}\label{effHam}
\end{eqnarray}
where ${\bf \sigma}$ is the Pauli matrix. $S_{\alpha\beta}({\bf q})$ 
and $C_{\alpha\beta}({\bf q})$ are 
the scattering matrices in which the particle-hole pair is scattered 
from the $d_{\alpha}$-orbital 
into the $d_{\beta}$-orbital, while $\bar{S}_{\alpha\beta}({\bf q})$ and 
$\bar{C}_{\alpha\beta}({\bf q})$ describe 
the scattering matrices in which 
$d_{\alpha}$-particle and $d_{\beta}$-hole is scattered into 
$d_{\beta}$-particle and $d_{\alpha}$-hole, respectively. 
These four scattering matrices 
are given as follows by treating the degenerate Hubbard model within RPA; 
\newpage
\begin{eqnarray}
  S_{\alpha\beta}({\bf q})&=&-U^{s}_{\alpha\beta}
  +\sum_{\gamma}(\hat{U}^{s}\hat{\chi}_{0}({\bf q}))_{\alpha\gamma}
  S_{\gamma\beta}({\bf q}),\nonumber\\
  C_{\alpha\beta}({\bf q})&=&-U^{c}_{\alpha\beta}
  -\sum_{\gamma}(\hat{U}^{c}\hat{\chi}_{0}({\bf q}))_{\alpha\gamma}
  C_{\gamma\beta}({\bf q}),\nonumber\\
  \bar{S}_{\alpha\beta}({\bf q})&=&\frac{-J'}
  {[1-(U'-J/2)\chi^{(0)}_{\alpha\beta}({\bf q})]^{2}
  -[J'\chi^{(0)}_{\alpha\beta}({\bf q})]^{2}},\nonumber\\
  \bar{C}_{\alpha\beta}({\bf q})&=&\frac{-J'}
  {[1-(U'+3J/2)\chi^{(0)}_{\alpha\beta}({\bf q})]^{2}
  -[J'\chi^{(0)}_{\alpha\beta}({\bf q})]^{2}}\eqnum{3}\label{scatmat}
\end{eqnarray}
where 
\begin{eqnarray}
  &&\hat{U}^{s}
 =\left[\begin{array}{ccc}
   U & -J & -J \\
   -J & U & -J \\
   -J & -J & U
        \end{array}\right],
    \hat{U}^{c}
 =\left[\begin{array}{ccc}
   U & 2U' & 2U' \\
   2U' & U & 2U' \\
   2U' & 2U' & U
        \end{array}\right],\nonumber\\
  &&\hspace*{5mm}\hat{\chi}_{0}({\bf q})
 =\left[\begin{array}{ccc}
   \chi^{(0)}_{xy}({\bf q}) & 0 & 0 \\
   0 & \chi^{(0)}_{yz}({\bf q}) & 0 \\
   0 & 0 & \chi^{(0)}_{zx}({\bf q})
        \end{array}\right],\nonumber\\
  &&\chi^{(0)}_{\alpha\beta}({\bf q})
  =\sum_{{{\bf k}}}
  \frac{f(\epsilon_{{\bf k}\alpha})-f(\epsilon_{{\bf k}+{\bf q}\beta})}
  {\epsilon_{{\bf k}+{\bf q}\beta}-\epsilon_{{\bf k}\alpha}}, 
  \hspace*{2mm}
  \chi^{(0)}_{\alpha}({\bf q})\equiv\chi^{(0)}_{\alpha\alpha}({\bf q})
  \eqnum{4}\label{intchi0}
\end{eqnarray}
where $f(\epsilon)$ is the Fermi distribution function. 
$S_{\alpha\alpha}({\bf q})$ and $C_{\alpha\alpha}({\bf q})$ are connected 
with the spin and charge fluctuations $for$ $d_{\alpha}$-$orbital$, 
respectively. 

The magnetic (charge or orbital) instability appears 
when the following relation is satisfied;
\begin{equation}
  {\rm{det}}[{\bf 1}\mp\hat{U}^{s(c)}\hat{\chi}_{0}({\bf q})]=0
  \hspace*{5mm}\rm{for}\hspace*{2mm}S_{\alpha\beta}({\bf q})
  (C_{\alpha\beta}({\bf q})).\eqnum{5}\label{instability}
\end{equation}
This equation of the instability for $S_{\alpha\beta}({\bf q})$ 
($C_{\alpha\beta}({\bf q})$) is reduced to the 
following cubic equation with respect to $J\hspace*{1mm}(2U')$;
\begin{eqnarray}
  &&x^{3}+a^{s(c)}x^{2}+c^{s(c)}=0, 
  \hspace*{5mm}x=J_{{\bf q}}\hspace*{1mm}(2U'_{{\bf q}})
  \nonumber\\
  &&a^{s(c)}=-\frac{1}{2}\sum_{\alpha}\frac{1}
  {\tilde{\chi}_{\alpha}^{s(c)}({\bf q})}, 
  \hspace*{5mm}
  c^{s(c)}=\frac{1}{2}\prod_{\alpha}\frac{1}
  {\tilde{\chi}_{\alpha}^{s(c)}({\bf q})}, 
  \nonumber\\
  &&\frac{1}{\tilde{\chi}_{\alpha}^{s(c)}({\bf q})}=
  \frac{1}{\chi^{(0)}_{\alpha}({\bf q})}{\mp}U\eqnum{6}\label{cubiceq}
\end{eqnarray}
where $J_{{\bf q}}$ ($U'_{{\bf q}}$) is the solution of the above 
cubic equation for each ${\bf q}$. 
When $\tilde{\chi}_{\alpha}^{s(c)}({\bf q})>0$ as expected, 
it can be proved that the above cubic equation gives always three real 
solutions (one negative and two positive). 
Since we have $J<0$ and $U'>0$, we should assign the negative solution 
of the upper one of eq. \ (\ref{instability}) to $J_{{\bf q}}$ 
and the smaller one of the positive solutions of the lower one of 
eq. \ (\ref{instability}) to $2U'_{{\bf q}}$. 
From this cubic equation, 
it is always expected that the absolute value of the solution for $J_{{\bf q}}$ 
decreases with the increase of $U$, while the value of the solution 
for $2U'_{{\bf q}}$ increase with $U$. Thus, $S_{\alpha\beta}({\bf q})$ 
is enhanced with the increase of $U$ or $|J|$, while $C_{\alpha\beta}({\bf q})$ 
developes due to the increase of $U'$. 
Also, it is expected that both instabilities of 
$S_{\alpha\beta}({\bf q})$ and $C_{\alpha\beta}({\bf q})$ 
should occur only within small regions in the ${\bf q}$-space 
where $\chi^{(0)}_{\alpha}({\bf q})$ has the peak structures. 
Instabilities for $\bar{S}_{\alpha\beta}({\bf q})$ and 
$\bar{C}_{\alpha\beta}({\bf q})$ are not found for realistic parameter values 
in the present case because of 
the small value of $\chi^{(0)}_{\alpha\beta}({\bf q})$ ($\alpha\neq\beta$). 

\begin{figure}[t]
  \centerline{\epsfxsize=7.5truecm \epsfbox{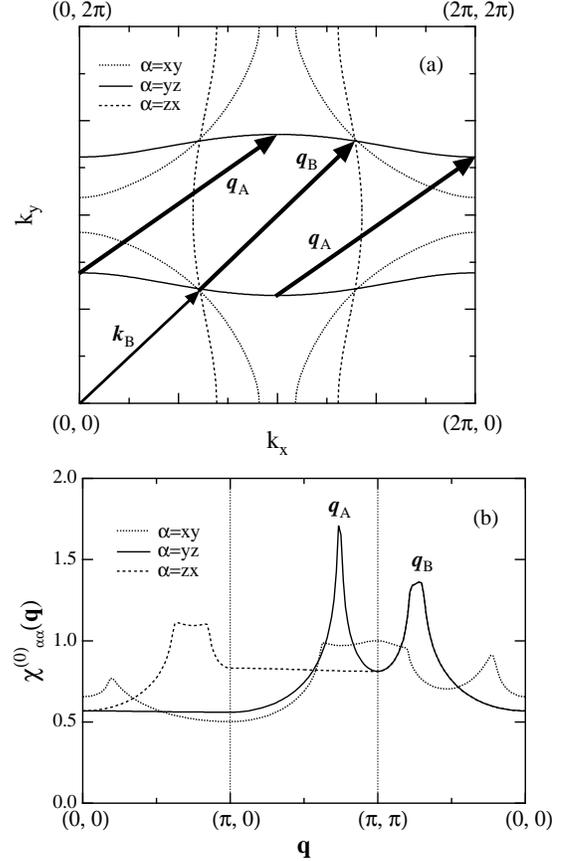} }
  \label{fig1}
  \caption
  {Calculated results of (a) Fermi surfaces and (b) 
  ${\bf q}$-dependences of 
  $\chi^{(0)}_{\alpha\alpha}({\bf q})$ using the parameter set 
  described in the text. ${\bf q}_{A}$ and ${\bf q}_{B}$ are 
  nesting vectors.}
\end{figure}

Carrying out the mean field prescription with respect to the obtained 
effective Hamiltonian, we derive the gap equation 
for the superconducting transition with the assumption that 
the effect of 
$\langle a_{{\bf k}\alpha\mu}^{\dagger}a_{-{\bf k}\beta\nu}^{\dagger}\rangle$ 
for $\alpha\neq\beta$ is negligibly small. 
This assumption seems to be reasonable, since the non-zero value of 
$\langle a_{{\bf k}\alpha\mu}^{\dagger}a_{-{\bf k}\beta\nu}^{\dagger}\rangle$ 
for $\alpha\neq\beta$ can be important only at the wave vector $\bf k$ 
where two Fermi surfaces cross with each other. We thus obtain the 
following gap equation for the singlet ($\eta=s$) and triplet ($\eta=t$) 
spin state;
\begin{eqnarray}
  \Delta^{\eta}_{\alpha}({\bf k})&=&
  \sum_{\beta}\sum_{{\bf k'}}V^{\eta}_{\alpha\beta}({\bf k}-{\bf k'})
  \frac{\tanh{(\epsilon_{{\bf k'}\beta}/2k_{\rm B}T)}}
  {2\epsilon_{{\bf k'}\beta}}
  \Delta^{\eta}_{\beta}({\bf k'})\nonumber\\
  &&V^{\eta}_{\alpha\alpha}({\bf q})=c_{\eta}S_{\alpha\alpha}({\bf q})
  +\frac{1}{2}C_{\alpha\alpha}({\bf q})\nonumber\\
  &&V^{\eta}_{\alpha\beta}({\bf q})=c_{\eta}\bar{S}_{\alpha\beta}({\bf q})
  +\frac{1}{2}\bar{C}_{\alpha\beta}({\bf q})\hspace*{5mm}(\alpha\neq\beta)
  \nonumber\\
  &&c_{s}=\frac{3}{2},\hspace*{5mm}c_{t}=-\frac{1}{2}\eqnum{7}\label{gapeq}
\end{eqnarray}
where $k_{\rm B}$ is the Boltzmann constant. 
We note that the superconducting transition occurs simultaneously at a 
temperature for all orbitals as far as we keep $J'\neq0$ 
or non-vanishing values of 
$\bar{S}_{\alpha\beta}({\bf q})$ and $\bar{C}_{\alpha\beta}({\bf q})$, 
as pointed out earlier by Agterberg $et$ $al.$.\cite{rf:17}

\begin{figure}[t]
  \centerline{\epsfxsize=7.5truecm \epsfbox{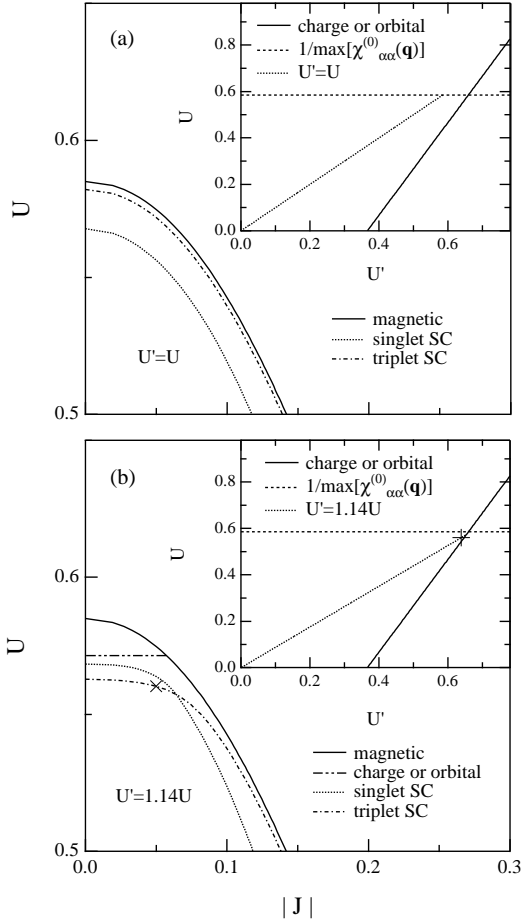} }
  \label{fig2}
  \caption
  {Calculated phase diagrams on $U$-$|J|$ plane 
  (a) for $U'=U$ and (b) for $U'=1.14U$. $|J|=J'$ is assumed. 
  Inset: Calculated phase diagrams on $U$-$U'$ plane. }
\end{figure}

Hereafter, we restrict ourselves to the small $|J|$ region with $J'=|J|$, 
the relation satisfied practically in any transition metals. \cite{rf:14}
We carried out the calculation at a fixed temperature 
$T=0.005$eV. Although this temperature looks too high compared 
with the observed value in the ruthenate, 
it seems to be sufficient insofar as we look for 
the mechanism of the superconductivity within the weak coupling theory 
without energy cut-off 
which overestimates the value of $T_{\rm c}$. 
When a solution is found at this temperature, lower values of $T_{\rm c}$ 
may be obtained easily by changing parameter values. 
In the parameter space around both instabilities for 
$S_{\alpha\beta}({\bf q})$ and $C_{\alpha\beta}({\bf q})$, 
we solve the gap equation for both the singlet and triplet spin states 
numerically, using a finite lattice with $128\times128$ meshes. 

\begin{figure}[t]
  \centerline{\epsfxsize=7.5truecm \epsfbox{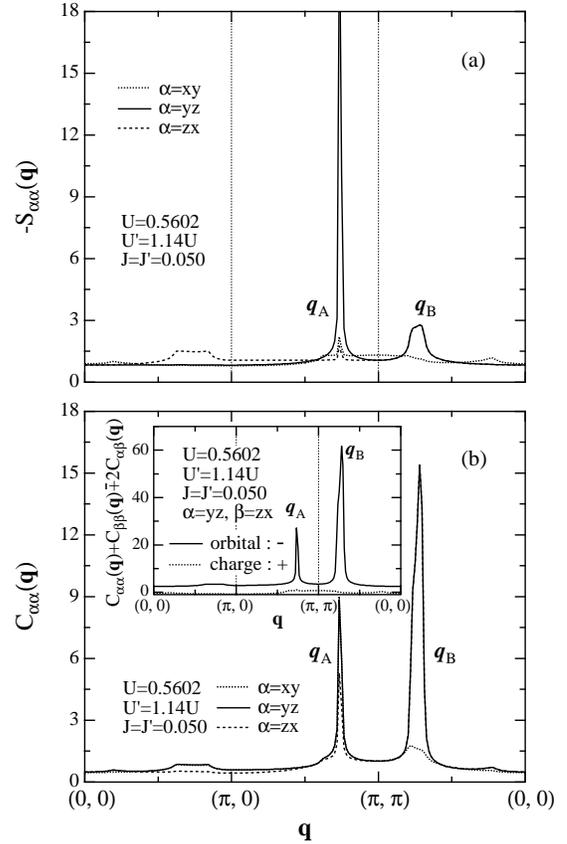} }
  \label{fig3}
  \caption
  {Calculated ${\bf q}$-dependences of 
  (a) $S_{\alpha\alpha}({\bf q})$ and (b) $C_{\alpha\alpha}({\bf q})$ 
  at a crossed point in Fig. 2(b). Orbital and charge component of 
  $C_{\alpha\alpha}({\bf q})$ ($\alpha=yz,zx$) are shown in the inset of 
  Fig. 3(b).}
\end{figure}

In order to decide the phase diagram at fixed temperature, 
we have to find out, first of all, the instabilities of 
$S_{\alpha\beta}({\bf q})$ and $C_{\alpha\beta}({\bf q})$. 
The instability of $S_{\alpha\beta}({\bf q})$ 
($C_{\alpha\beta}({\bf q})$)
is given by $|J_{\rm cri}|={\rm min}|J_{{\bf q}}|$ 
($U'_{\rm cri}={\rm min}U'_{{\bf q}}$). 
We show the phase diagrams of this model for $U'=U$ and 
$U'=1.14U$ in Fig. 2(a) and 2(b), respectively. The mode of 
the magnetic state in Fig. 2 is ${\bf q}_{\rm A}$ shown in Fig. 1. 
Thus, it is easily understood that the singlet superconductivity in Fig. 2(a) 
has the symmetry of $d_{x^2-y^2}$ due to the enhanced 
spin fluctuations around the antiferromagnetic mode \cite{rf:12}. 
On the other hand, it is seen in Fig. 2(b) that the instability line of 
$C_{\alpha\beta}({\bf q})$ at ${\bf q}={\bf q}_{\rm B}$ gets into 
the paramagnetic parameter space in the small $|J|$ region. 
We find that both 
$S_{\alpha\alpha}({\bf q})$ and $C_{\alpha\alpha}({\bf q})$ have 
the peak structures at the same ${\bf q}$-points as shown in Fig. 3. 
Thus, from the gap equation we see that 
the triplet pairing interaction is enhanced due to the strong 
and relatively wide peak structure of $C_{\alpha\alpha}({\bf q})$ 
($\alpha=yz,\hspace{1mm}zx$) around ${\bf q}={\bf q}_{\rm B}$ 
while the singlet one is considerably reduced with the increase of $U'$. 
Thus, increasing the value of $U'$ favors the triplet superconducting state 
rather than the singlet one. 
In order to clarify what contributes to $C_{\alpha\alpha}({\bf q})$, 
we define the orbital (charge) operators 
from $d_{yz}$- and $d_{zx}$-orbitals as 
$n_{iyz} \mp n_{izx}$, 
and show in the inset of Fig. 3(b) the ${\bf q}$-dependences of 
$C_{\alpha\alpha}({\bf q})+C_{\beta\beta}({\bf q}) \mp 
2C_{\alpha\beta}({\bf q})\hspace{1mm}(\alpha=yz,\hspace{1mm}\beta=zx)$ 
where 'minus' and 'plus' are connected with the orbital and charge 
fluctuations, respectively. It is evident that 
the orbital fluctuation between $d_{yz}$- and $d_{zx}$-orbitals 
mostly contributes to the ${\bf q}$-dependences of 
$C_{\alpha\alpha}({\bf q})$ ($\alpha=yz,\hspace{1mm}zx$). 
Also, we show in Fig. 4 the ${\bf k}$-dependence of 
$\Delta^{t(y)}_{yz}({\bf k})$ where $\Delta^{t(\xi)}_{\alpha}({\bf k})$ 
is odd with respect to $k_{\xi}$. 
Under this condition the magnitude of $\Delta^{t(y)}_{yz}({\bf k})$ is 
much larger than $\Delta^{t(y)}_{zx}({\bf k})$ 
and $\Delta^{t(y)}_{xy}({\bf k})$. 
From this figure, it is seen that the triplet superconducting state 
in Fig. 2(b) belongs to the $A_{1g}{\times}E_{u}$ irreducible representation 
under tetragonal symmetry $D_{4h}$, because 
$\Delta^{t(y)}_{yz}({\bf k})$ at $k_{x}=0$ has the non-zero value 
and the direction of the node deviates from the diagonal. 
We may think of the form of the gap function 
$\Delta^{t(y)}_{yz}({\bf k})$ proportional to 
$\sin{k_{y}}(A+B(\cos{k_{x}}+\cos{k_{y}}))$. We show 
in Fig. 4 a rough fitting result due to this function. 
Non-zero value of $A$ seems to be essential for better fitting. 
For comparison, we also show a fitting by using the function 
$\sin{k_{y}}(\cos{k_{x}}-\cos{k_{y}})$(belonging to $B_{1g}{\times}E_{u}$). 
Thus, summarizing these results, it is understood that the 
superconducting phase obtained in Fig. 2(b) is the phase of 
the orbital dependent triplet superconducting state 
which is induced by the strong orbital fluctuations 
between $d_{yz}$- and $d_{zx}$-orbital 
with the increase of $U'$. However, it remains 
to discuss if the value of the parameter $U'$ used in Fig. 2(b) 
is realistic. 

It seems difficult to explain the larger value of $U'$ compared with $U$ 
insofar as we use the bare values for them. In actuality we should use 
$renormalized$ values for these parameters in the present RPA scheme. 
In order to discuss the values of the renormalized interactions, 
we extend the multiple scattering problem \cite{rf:14} between two electrons 
to the present case. For simplicity, we assume $|J|=J'=0$, 
since we are interested only in the small $|J|$ region. 
As usual, the renormalized interaction strengthes 
$\bar{U}_{\alpha}, \bar{U'}_{\alpha\beta}$
with the momentum ${\bf q'}$ for the center of gravity of two particles 
are given as follows;
\begin{eqnarray}
  &&\bar{U}_{\alpha}({\bf q'})
 =\frac{U_{bare}}{1+U_{bare}\phi^{(0)}_{\alpha\alpha}({\bf q'})},\nonumber\\
  &&\bar{U'}_{\alpha\beta}({\bf q'})
 =\frac{U'_{bare}}{1+U'_{bare}\phi^{(0)}_{\alpha\beta}({\bf q'})}
 \hspace*{5mm}(\alpha\neq\beta),
 \nonumber\\
  &&\phi^{(0)}_{\alpha\beta}({\bf q'})
  =\sum_{{{\bf k}}}
  \frac{1-f(\epsilon_{-{\bf k}\alpha})-f(\epsilon_{{\bf k}+{\bf q'}\beta})}
  {\epsilon_{-{\bf k}\alpha}+\epsilon_{{\bf k}+{\bf q'}\beta}}, 
  \eqnum{8}\label{tmatrix}
\end{eqnarray}
where $U_{bare}$ and $U'_{bare}$ are the $bare$ interaction constants 
which satisfy the usually expected relation $U'_{bare}=U_{bare}$. 
Now, we consider the case in which a particle at ${\bf k}+{\bf q}_{B}$ 
and a hole at ${\bf k}$ are scattered to a particle at ${\bf k'}+{\bf q}_{B}$ 
and a hole at ${\bf k'}$ by $\bar{U}_{\alpha}$ or $\bar{U'}_{\alpha\beta}$. 
Because the value of $\chi^{(0)}_{yz(zx)}({\bf q}_{B})$ 
is principally attributed to the contribution around ${\bf k}={\bf k}_{B}$ 
shown in Fig. 1(a), we have ${\bf k}\approx{\bf k'}\approx{\bf k}_{B}$, 
that is, ${\bf q'}={\bf k}+{\bf k'}+{\bf q}_{B}\approx(2\pi,2\pi)$. 
It is expected that $\phi^{(0)}_{\alpha\alpha}({\bf q'})$ is much larger than 
$\phi^{(0)}_{\alpha\beta}({\bf q'})$ ($\alpha\neq\beta$) around 
${\bf q'}={\bf 0}$ because of the different geometry between these two 
Fermi surfaces. As a result, $\bar{U}_{\alpha}({\bf q'})$ is more suppressed 
than $\bar{U'}_{\alpha\beta}({\bf q'})$ around ${\bf q'}={\bf 0}$. Thus, 
if $U$ and $U'$ are regarded as the renormalized interactions, 
$U'>U$ seems to be satisfied in our case. 
This result seems to favor the present mechanism for 
the orbital instability as shown in Fig. 2(b). 
For more quantitative estimation, we need more detailed studies in future. 

\begin{figure}[t]
  \centerline{\epsfxsize=7.5truecm \epsfbox{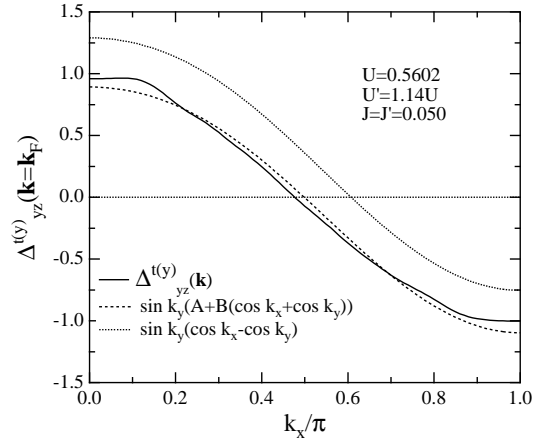} }
  \label{fig4}
  \caption
  {Calculated ${\bf k}$-dependences of $\Delta^{t(y)}_{yz}({\bf k})$ 
  along the $d_{yz}$ Fermi surface in the first quadrant. 
  Selected parameter set is the same as Fig. 3. 
  Fitting parameters are chosen as $A=0.51$, $B=1.3$.}
\end{figure}

It is interesting to note that even though 
the calculated intensity of the spin fluctuations as shown in 
Fig. 3(a) has the peaks around the antiferromagnetic mode in accordance 
with recent inelastic neutron experimental results \cite{rf:11}, 
we obtain the triplet 
superconducting phase rather than singlet one in a fairly plausible 
parameter regime. It is evident that the strong orbital fluctuations enhanced 
by the increase of $U'$ as compared with $U$ 
play the most important role in this mechanism. 
We expect the present mechanism to apply to other strongly correlated 
electron systems with degenerate orbitals and multiple Fermi surfaces. 
In particular, it is tempting to consider a possible application 
of this mechanism to $\rm{UPt}_{3}$ 
which is another triplet superconductor with both the multiple Fermi 
surfaces and the strong antiferromagnetic correration.  

In conclusion, 
we have investigated 
the mechanism of superconductivity in ${\rm Sr}_{2}{\rm RuO}_{4}$ 
using the Hubbard model with degenerate orbitals within the weak coupling 
theory. It is shown that the triplet superconductivity appears, 
when the strongly correlated system approaches the orbital instability 
achieved by the increase of the on-site Coulomb interaction $U'$ 
between the electrons in different orbitals as compared with $U$
between the electrons in the same orbital. 
It should be stressed that this superconducting mechanism is only available 
in the orbitally degenerate system with multiple Fermi surfaces.

This author is indebted Prof. T. Moriya, Prof. K. Ueda, Prof. K. Yamaji 
and Prof. T. Yanagisawa for useful discussions. 
The author is suported by 
Japan Science and Technology Corporation, 
Domestic Research Fellow.

\end{document}